\documentclass[lettersize,journal]{IEEEtran}
\usepackage{amsmath,amsfonts}
\usepackage{algorithmic}
\usepackage{algorithm}
\usepackage{amsfonts}
\usepackage{array}
\usepackage{amsmath}
\usepackage[caption=false,font=normalsize,labelfont=sf,textfont=sf]{subfig}
\usepackage{textcomp}
\usepackage{stfloats}
\usepackage{url}
\usepackage{verbatim}
\usepackage{graphicx}
\usepackage{cite}
\usepackage[utf8]{inputenc}
\usepackage{amssymb}
\usepackage{bm}
\usepackage{multirow}
\usepackage{braket}
\newcommand{\sign}[0]{\ensuremath{\text{sign}}}
\usepackage{xcolor}

\begin{document}
\title{AutoQML: Automatic Generation and Training of Robust Quantum-Inspired Classifiers by Using Genetic Algorithms on Grayscale Images}

\author{Sergio Altares-López, Juan José García-Ripoll, Angela Ribeiro
\thanks{

Sergio Altares-López is with Consejo Superior de Investigaciones Científicas, Centre for Automation and Robotics (CAR),CSIC-UPM, Ctra. Campo Real km. 0,200, 28500 Arganda, Spain and Universidad Politécnica de Madrid, Programa de Doctorado en Automática y Robótica, Calle de José Gutiérrez Abascal 2, 28006 Madrid, Spain. (e-mail: sergio.altares@csic.es).

Juan José García-Ripoll is with Consejo Superior de Investigaciones Científicas, Instituto de Física Fundamental (IFF), CSIC, Calle Serrano 113b, 28006 Madrid, Spain. (e-mail: juanjose.ripoll@csic.es).

Angela Ribeiro is with the Consejo Superior de Investigaciones Científicas, Centre for Automation and Robotics (CAR), CSIC-UPM, Ctra. Campo Real km. 0,200, 28500 Arganda, Spain. (e-mail: angela.ribeiro@csic.es).

The authors gratefully acknowledge the computer resources at Artemisa, funding from the European Union ERDF and Comunitat Valenciana, and technical support provided by the Instituto de Física Corpuscular, IFIC (CSIC-UV). Work partially funded by Ministerio de Asuntos Económicos y Transformación Digital, project - MIA.2021.M01.0004.}} 

\markboth{}%
{Altares-López \MakeLowercase{\textit{et al.}}: AutoQML: Automatic Generation and Training of Robust Quantum-Inspired Classifiers by Using Genetic Algorithms on Grayscale Images}

\maketitle

\begin{abstract}
We propose a new hybrid system for automatically generating and training quantum-inspired classifiers on grayscale images by using multiobjective genetic algorithms. We define a dynamic fitness function to obtain the smallest possible circuit and highest accuracy on unseen data, ensuring that the proposed technique is generalizable and robust. We minimize the complexity of the generated circuits in terms of the number of entanglement gates by penalizing their appearance. We reduce the size of the images with two dimensionality reduction approaches: principal component analysis (PCA), which is encoded in the individual for optimization purpose, and a small convolutional autoencoder (CAE). These two methods are compared with one another and with a classical nonlinear approach to understand their behaviors and to ensure that the classification ability is due to the quantum circuit and not the preprocessing technique used for dimensionality reduction.
\end{abstract}


\section{Introduction}

\IEEEPARstart{T}{he} field of quantum computing is currently evolving, developing advantages over classical computing. This new computation paradigm is based on the ability to use physical properties such as \textit{entanglement} or \textit{superposition}, which allow \textit{quantum bits} or \textit{qubits}, which are basic information units in quantum computing, to be in more than one state at the same time (see Section ~\ref{subsec:qubits}), allowing access to Hilbert spaces and thus to spaces that may be infinite-dimensional $\mathcal{H}$. Similar to classical computing, in which information is calculated based on electrical circuits and logic gates that operate on bits, in quantum computing, \textit{quantum circuits} composed of sequences of \textit{quantum gates} are used to modify quantum states (see Section ~\ref{subsec:qugates}).

In the era of \textit{big data}, due to the large amount of information generated, the computational speed of data processing is a limiting factor for classical computing. Quantum computing can exponentially increase the computational capacity by using certain physical properties. Thus, a classical computer would need $2^{200}$ bits to process the same amount of information represented by 200 qubits ~\cite{nielsen}. Additionally, the use of quantum computing allows a larger memory capacity with fewer resources than a classical machine, resulting in cost savings. These benefits can be used in a variety of fields, such as insurance for portfolio optimization, route optimization for logistics, cybersecurity through quantum random number generation for cryptography or the development of new materials and drugs.
\textit{Quantum machine learning} (QML)~\cite{Biamonte_2017,Schuld_2018,Buffoni_2020} is a new and rapidly evolving paradigm based on the combination of \textit{machine learning} (ML) and quantum computing. QML explores the potential advantages of operating in a Hilbert space and using quantum operations for common tasks in both supervised and unsupervised learning with both classical and quantum data representations. Early studies explored the advantages of quantum computing for classification~\cite{riste2017demonstration}, model generalization~\cite{gao2018quantum} and model training~\cite{winci2020path}. Some of the proposed advantages of QML include access to larger feature spaces, more general and expressive models and algorithmic improvements in model optimization. Our proposed quantum image classification methods utilize these advantages. We optimize the models by using metaheuristic techniques to automatically obtain the best predictions based on test data, ensuring model robustness, and we automatically generate simple quantum-inspired machine learning classifiers that can easily be implemented on classical computers.

In addition to algorithms that are useful for future scalable fault-tolerant quantum computers \cite{HHL17}, the field of QML investigates the utility of near-term intermediate-scale quantum (NISQ) devices, which are usually based on \textit{parameterized quantum circuits} (PQC)~\cite{NISQ18,NISQ20,NISQ21,NISQ22,benedetti2019parameterized,par20,du2020expressive,cao2020cost,cerezo2021cost} (see section~\ref{subsec:qugates}). There has been substantial progress in the development of algorithms for different tasks, such as classification with quantum support vector machines~\cite{rebentrost2014quantum, Havlcek2019}, quantum autoencoders for feature extraction ~\cite{huang2020realizationae,bravo2021quantumae,cao2021noiseae,srikumar2021clusteringae}, quantum generative adversarial neural networks (QGANs) ~\cite{huang2021experimental,herr2021anomaly,niu2021entangling} and quantum neural networks (QNNs), which use measurement probabilities for predictions ~\cite{abbas2021powernn,farhi2018classificationnn,killoran2019continuousnn}.

The most common methods for obtaining these PQCs are to generate them manually or to use predefined general templates that do not incorporate any understanding of the datasets or use cases. Several methods have been proposed for partially optimizing parameters or circuit structures 
by using techniques such as adaptive pruning or reinforcement learning ~\cite{sim2021adaptive, ostaszewski2021structure,ostaszewski2021reinforcement,bilkis2021semi}.Also, we find studies which optimize both parameters and circuit structure at the same time by using genetic algorithms \cite{2021saj}. Furthermore, other studies have used different techniques, including variational autoencoders for PQC optimization ~\cite{chen2022generating} and evolutionary algorithms for parameter optimization ~\cite{anand2021naturalgen}. Some studies also have used evolutionary techniques to generate quantum autoencoders ~\cite{lamata2018quantugenm}. Another research direction is to reduce the number of entangling operations in PQCs, thus reducing their complexity ~\cite{compact2022}. This concept is important because entanglement gates increase the computational cost of the circuit.

Deep learning techniques have classically been used for image classification tasks, and the training process has a high computational cost. To address this problem by using quantum artificial intelligence, we must consider the high dimensionality of these data when adding the feature vector to the quantum circuit. There are several approaches for creating quantum convolutional neural networks ~\cite{qcnn2019quantum,wei2022quantum_conv}, which are hybrid classical-quantum systems that use classical convolutional networks ~\cite{liu2021hybrid}, as well as interesting approaches such as making \textit{transfer learning} from pretrained classical models to quantum neural networks ~\cite{transf2020}.

The application of QML to high-dimensional data is currently a challenge. Previously developed models normally use a classical component, such as deep learning or multilayer perceptron (MLP), to extract features to obtain the final classification. 
Due to their nature, deep learning systems considerably reduce the dimensionality of the data, providing, depending on data, a direct linear relationship between the data and the target. Thus, the quantum circuit is not predictive, and the interpretability of the output is lost. Another point to consider is that because deep learning systems rely on computationally expensive pretraining that is specific to $N$ categories, they are specialist systems that do not allow a fast use.

As we have seen, the design of quantum algorithms for high-dimensional data is complex, expensive computationally, and usually involves hybrid systems.

In this work, we extend the strategy \cite{2021saj}, in which we generate quantum kernels genetically for small tabular datasets using a basic parameter and gate encoding, which yielded satisfactory results. We apply this strategy to a higher dimensional dataset ---images, in order to evaluate the maximum performance that can be achieved by using this method. We also improve the encoding in terms of gate types and parameters, increasing the flexibility of the system. 

Similar to how \textit{evolutionary neural networks} generate trained networks with optimized parameters, biases and structures by applying genetic algorithms to achieve the best accuracy ~\cite{liu2021gann,leung2003gann, shojaeefard2012gann, suganuma2017gann}, we propose a new system that uses evolutionary techniques to automatically generate and train quantum classifiers on grayscale images. Our system is based on a nontrainable dimensionality reduction method that is optimized in the system itself since the technique is incorporated into the individual encoding (see Section ~\ref{sec:DR}). Furthermore, we implement an alternative \textit{transfer learning} method ~\cite{transfer1,transfer,transf2020} based on a small convolutional autoencoder network to reduce the dimension and analyze the classifier behavior. Thus, since we created a nonlinear MLP with a size and number of trainable parameters comparable to a quantum system as a fair prediction baseline, we guarantee that the classification power is due to the generated circuit.

In this study, we determine not only the best circuit topology but also the optimal parameters by using a multiobjective genetic algorithm to achieve the best accuracy and smallest circuit size, thus reducing the expressivity and complexity of the system. All the investigations described in this work were calculated on a kernel matrix in a Python simulator of an ideal quantum computer ~\cite{2021saj}.

This work is formally introduced in Section \ref{sec:QSVM}, in which we describe the main quantum computing elements used in QML, support vector machines and quantum support vector machines. Then, the evolutionary quantum-inspired classifier technique for grayscale images is described in Section \ref{sec:EQC}, including its encoding, fitness function and genetic operators. Then, the datasets used in this study are described in Section \ref{sec:DS}. We describe the dimensional reduction methods used due to the image size in Section \ref{sec:DR}. We introduce the training process in Section \ref{sec:AL}, including the main steps for implementing our new technique. Next, the results obtained for each of the datasets with each of the techniques are presented in Section \ref{sec:Reesults}. Finally, the conclusions and future research directions are presented in Section \ref{sec:conclusions}. We also include an appendix \ref{sec_ap}, which discusses details on the gates and angle binary encoding.

\section{Quantum Methods for Supervised Learning}
\label{sec:QSVM}
This section introduces the basic concepts of quantum computing, including qubits, the quantum operators that change quantum states and quantum circuits.

\subsection{Qubits}
\label{subsec:qubits}
Analogous to the bit in classical computing, the \textit{qubit} is the smallest unit information that can be processed in a quantum computer. The set of possible states of a qubit forms a Hilbert space. This space has two distinguishable states that can be represented as
\begin{equation}
\ket{0} = \begin{pmatrix}
    1\\
    0 \\
  \end{pmatrix},\;
\ket{1} = \begin{pmatrix}
    0\\
    1 \\
  \end{pmatrix}
\end{equation}
in Dirac's notation. These states are analogous to the 0 and 1 states of a classical bit. However, qubits can also be found in arbitrary quantum \textit{superpositions} of these basis states. These superpositions are linear combinations of the basis states, $\ket{\psi} = \psi_0{\ket{0} + \psi_1\ket{1}}$, with normalized complex amplitudes $|\psi_0|^2 + |\psi_1|^2 = 1$ that represent the probabilities $|\psi_0|^2$ and $|\psi_1|^2$ of detecting states $\ket{0}$ and $\ket{1}$ when measuring the qubit in the quantum computer.

General quantum states with two or more qubits are constructed as tensor product representations of the Hilbert spaces of the qubits. For instance, a general two-qubit state is a linear combination of the $2^2$ basis states $\ket{\psi} = \psi_{00}\ket{00} + \psi_{01}\ket{01}+ \psi_{10}\ket{10}+ \psi_{11}\ket{11}$, where $|\psi_{nm}|^2$ is the probability of obtaining the bits $n, m\in\{0,1\}$ when measuring the two qubits and $\sum_{mn}|\psi_{mn}|^2=1$. Due to the composition of Hilbert spaces, the number of complex coefficients used to represent a quantum state with $N$ qubits increases exponentially as $2^N$. Thus, a general purpose quantum computer with 40 or more qubits cannot be simulated in a supercomputer by classical means. However, this relation also suggests that we can use very large feature spaces to implement sophisticated machine learning models.

\subsection{Quantum Gates and Quantum Circuits}
\label{subsec:qugates}

The evolution of a perfect isolated quantum state is reversible and can be represented as a unitary operator. These operators are matrices $\mathcal{U}$ that transform a state vector into another complex state vector $\ket{\phi_{2}} = \mathcal{U}\ket{\phi_{1}}$ and whose adjoint is the inverse operation $\mathcal{U} \mathcal{U}^{\dagger} = I$.

An earlier quantum computing result demonstrated that arbitrary transformations of quantum states can be decomposed into smaller sets of universal gates~\cite{Intro, nielsen}, including all single-qubit operations and at least one two-qubit maximally entangling gate. These \textit{quantum logic gates} are the quantum equivalent of Boolean operations such as AND, OR, NOT, and XOR.

\begin{figure*}[!t]
\centering
\subfloat[]{\includegraphics[width=3.3in]{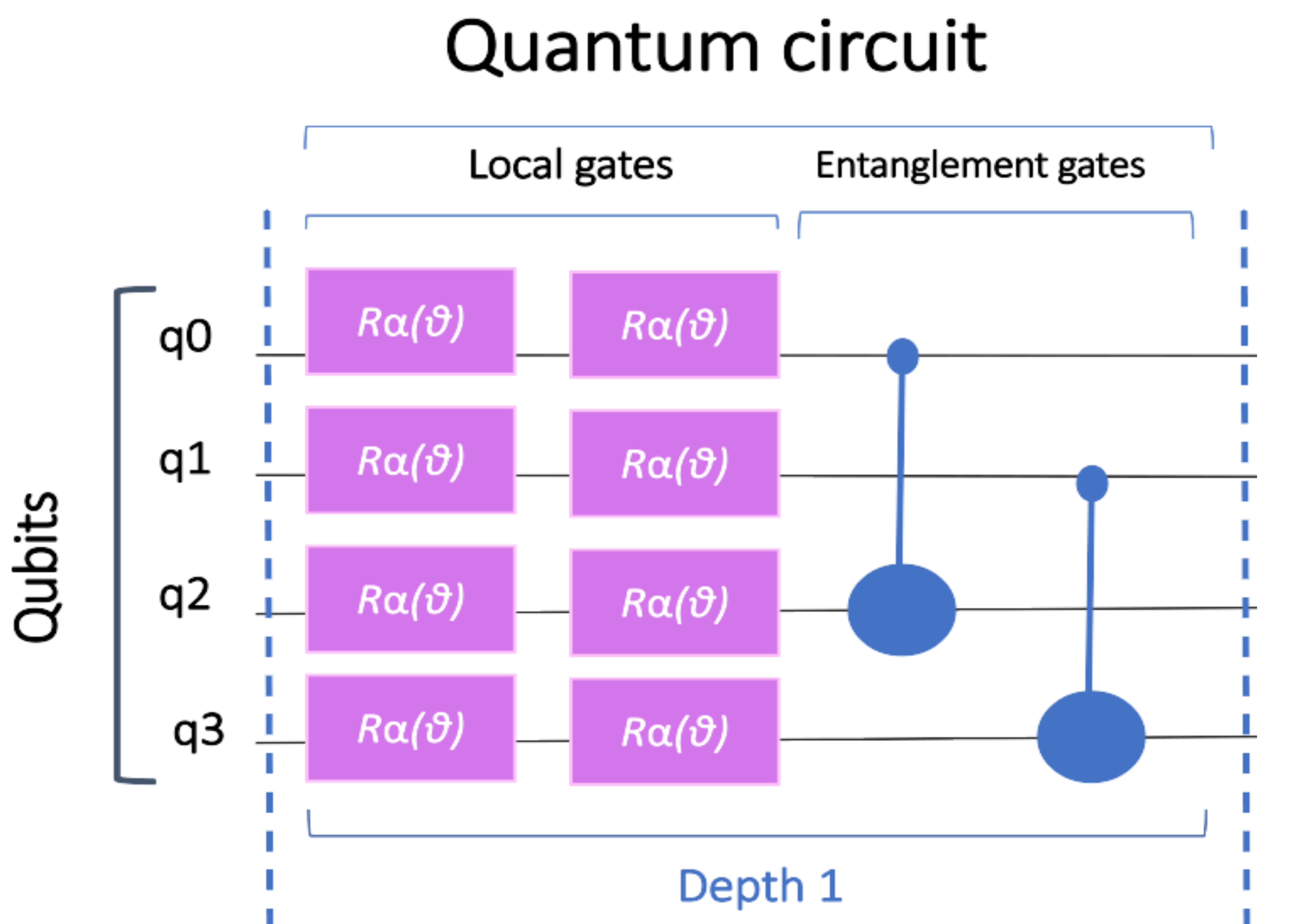}%
\label{fig:fig_first_case0}}
\hfill
\subfloat[]{\includegraphics[width=3.3in]{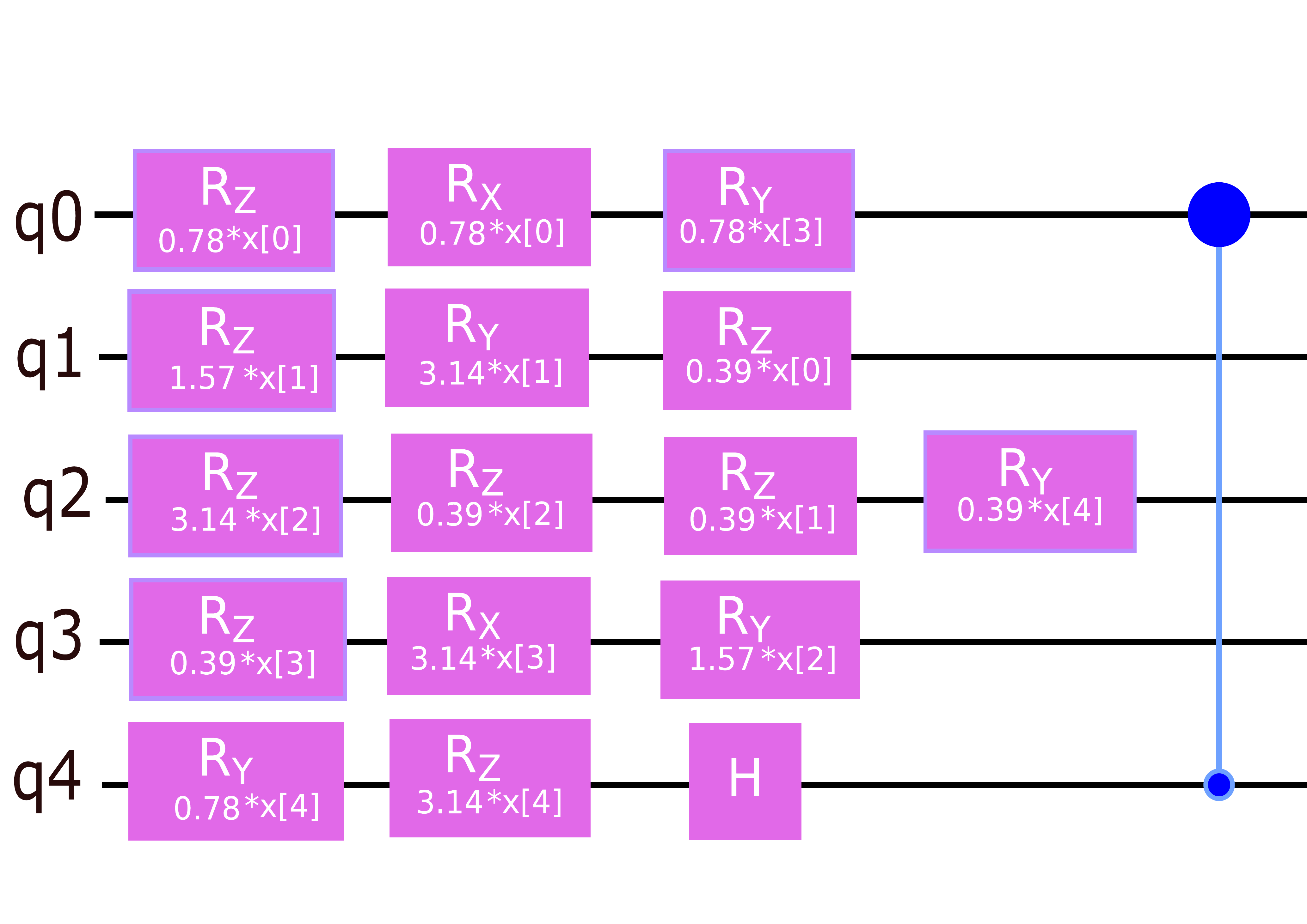}%
\label{fig:fig_second_case0}}

\caption{a) A general quantum circuit. We focus on the essential components of the quantum circuit. First, we consider the lines of qubits where the quantum gates are located. These operators can act locally on a single qubit, such as rotations, which normally depend on the parameter $\theta$, or they can act on several qubits, such as entanglement gates. Rotations ($R_{\alpha}$) can be performed on the X, Y and Z axes. The structure can be repeated N times. The number of times that the structure is repeated is called the depth of the circuit. b) We show a quantum circuit that was generated automatically in our previous work ~\cite{2021saj}. This circuit acts as the \textit{quantum feature map} for a drug classification dataset ~\cite{drug_cl} and consists of five qubits with parameterized input-dependent gates, including H-gates and an entanglement gate, in which the variables of the dataset are embedded.}
\label{fig_23}
\end{figure*}

A quantum algorithm can be represented as a \textit{quantum circuit}, which is a concatenation of quantum logic gates and measurements. When a circuit is sufficiently small, it can be drawn using a ``pentagram-like'' notation, as shown in Figure~\ref{fig_23}a. Each horizontal line denotes a qubit, subject to the quantum gates that are placed on top of those lines, or connecting them. 
 
In some instances, such as the problems discussed in this work, quantum circuits can depend on classical external data. Figure~\ref{fig_23}b shows a genetically optimized quantum circuit that is capable of classifying data in a standard QML framework~\cite{2021saj}.

In this work, we design quantum circuits using a slightly redundant set of quantum gates, single-qubit operations and one two-qubit gate. General single-qubit rotations are best expressed as three-dimensional rotations in the qubit's Bloch sphere space. These rotations are generated by the Pauli matrices:
\begin{equation}
\sigma_{x} =  \begin{pmatrix}
    0 & 1\\
     1 & 0\\
  \end{pmatrix},\,
\sigma_{y} =  \begin{pmatrix}
    0 & -i\\
     i & 0\\
  \end{pmatrix},\,\sigma_{z} =  \begin{pmatrix}
    1 & 0\\
     0 & -1\\
  \end{pmatrix}.
\end{equation}
A generic single-qubit gate then adopts a parameterized form:
\begin{equation}
  R_3(\theta_x,\theta_y,\theta_z) = 
 \exp\left(-\tfrac{i}{2}(\theta_x\sigma_x+\theta_y\sigma_y+\theta_z\sigma_z)\right).
  \label{eq:R3}
\end{equation}
This form is available in several quantum computing platforms. This parameterization includes the quantum equivalent of the NOT operation $X=\sigma_x$, which transforms the state of a qubit with a Pauli-X rotation:
\begin{equation}
X\ket{0} = \ket{1},\; X\ket{1} = \ket{0}.
\end{equation}
This parameterization also includes the \textit{Hadamard} gate, a gate without a classical analog that enables the creation of superposition states. This gate transforms the basis states $\ket{0}$ and $\ket{1}$ to $\tfrac{1}{\sqrt{2}}(\ket{0} +\ket{1})$ and $\tfrac{1}{\sqrt{2}}(\ket{0} - \ket{1})$, respectively, as follows:

\begin{equation}
H = \frac{1}{\sqrt{2}}
  \begin{pmatrix}
    1 & 1\\
    1 & -1\\
  \end{pmatrix}.
\end{equation}

To train the quantum circuits more simply, we introduce specialized rotations along each of the three axes (X, Y, Z):
\begin{eqnarray}
R_{x}(\theta) = R_3(\theta,0,0)
&=& \begin{pmatrix}
    \cos \frac{\theta}{2} & - i \sin \frac{\theta}{2}\\
     - i \sin \frac{\theta}{2} & \cos \frac{\theta}{2}\\
   \end{pmatrix},\notag\\
R_{y}(\theta) = R_3(0,\theta,0)
&=& \begin{pmatrix}
    \cos \frac{\theta}{2} & - \sin \frac{\theta}{2}\\
     \sin \frac{\theta}{2} & \cos \frac{\theta}{2}\\
  \end{pmatrix},\\
R_{z}(\theta) = R_3(0,0,\theta)
&=& \begin{pmatrix}
    e^{-i\frac{\theta}{2}} & 0\\
    0 & e^{i \frac{\theta}{2}}\\
  \end{pmatrix}.\notag
\end{eqnarray}
Note that a combination of these rotations is equivalent to the general formula presented in~\eqref{eq:R3}.

Finally, the quantum circuits need an entangling operation with at least two qubits. We chose the \textit{Controlled-NOT} (\textit{CNOT}) gate, an operation that is a very common primitive in most quantum computing platforms. This gate is a simple, maximally entangling gate that flips the state of the target qubit only when the control qubit is in the $1$ state. In matrix notation, this gate has the form of the expression~\ref{eq:cnot} such as,
\begin{equation}
CNOT = \begin{pmatrix}
    1 & 0&0&0\\
    0 & 1&0&0\\
    0 & 0&0&1\\
    0 & 0&1&0\\
  \end{pmatrix}
  \label{eq:cnot}.
\end{equation}

\subsection{Quantum Support Vector Machine (QSVM)}
\label{subsec:qsv}

In this study, we explore the application of quantum computers in the implementation of a classifier. For simplicity, we focus on a problem with two classes, in which we attach a label $y \in \{-1,+1\}$ to each feature vector $\textbf{x} = \{x_1,x_2 \dots x_N\} \in \mathbb{R}^N$. Our goal is to develop a quantum model $g(\textbf{x})$ that not only predicts the correct labels for the training set but also generalizes well to unseen test data.

In a classical setting, one of the simplest supervised learning algorithms is the \textit{support vector machine} (SVM). This technique classifies data using a hyperplane that divides the feature space into two sections, with each section assigned the label +1 or -1 according to
\begin{equation}
  g(\mathbf{x}) = \sign\left(\mathbf{w}^{T} \cdot \mathbf{x} + b\right).
  \label{eq:hyperplane}
\end{equation}
The normal $\textbf{w}$ and position $b$ of the hyperplane are derived according to \textit{support vectors} that are extracted from the data by maximizing the loss function ~\ref{eq:loss}, which measures the separation between the data and the hyperplane: 
\begin{equation}
  \mathbf{w} = \sum_i \gamma_i y_i \mathbf{x}_i,\,
  \label{eq:loss}
\end{equation}
The SVM algorithm can be generalized for nonlinearly separable data by introducing \textit{feature maps}. Feature maps are functions that map the data to be classified to a different vector space $\tilde{\mathbf{x}}_i := \bm{\Phi}(\mathbf{x}_i) \in \mathbb{R}^{r}$ where the data are linearly separable. According to Mercer's theorem~\cite{geron2019hands}, the generalized SVM does not need to know the feature map $\bm{\Phi}$, which may be an infinite dimensional function. Instead, the full classifier $g(\mathbf{x})$, can be recovered using only the \textit{kernel function} $K(\mathbf{x},\mathbf{x}')=\bm{\Phi}(\mathbf{x})^T \bm{\Phi}(\mathbf{x}')$ of the scalar products between points in the new feature space such as the expression~\eqref{eq:funcion_decision},
\begin{equation}
  \label{eq:funcion_decision}
  g(x) = \sign\left(\sum_i \gamma_i y_i K(\mathbf{x}_i,\mathbf{x}) + b \right) \in \{+1,-1\}.
\end{equation}

The \textit{quantum support vector machine} (QSVM)~\cite{rebentrost2014quantum,Li_2015,Havlcek2019,schuld2019} is a unifying framework for various classifier strategies that use quantum circuits with classical data. In this framework, each feature vector $\mathbf{x}$ is associated with a quantum state $\ket{\Phi(\mathbf{x})} := \mathcal{U}(\mathbf{x};\bm{\theta})\ket{0^n}$ that is created using a quantum circuit---the \textit{quantum feature map}---with some trainable parameters $\bm{\theta}$, as shown in Figure~\ref{fig_23}. Although the quantum state can be processed in different ways~\cite{Li_2015,Havlcek2019}, it has been shown that all methods are equivalent to a generalized SVM with a quantum kernel function~\cite{schuld2021quantumkernel}
\begin{eqnarray}
  K(\mathbf{x}, \mathbf{x}')
  & = &
  \mathrm{Re}\braket{\Phi(\mathbf{x})|\Phi(\mathbf{x}')}\\
  & = &
  \mathrm{Re}\braket{0^n | \mathcal{U}(\mathbf{x};\bm{\theta})^\dagger \mathcal{U}(\mathbf{x}';\bm{\theta})|0^n}.
  \label{eq:new-kernel}
\end{eqnarray}
The kernel function (\ref{eq:new-kernel})
 can be estimated numerically for smaller problems. However, in other situations, the kernel function must be approximated experimentally with a quantum computer.

This work explores the classification power of quantum SVMs in image processing applications. In accordance with our previous work, this study focuses on small parameterized circuits $\mathcal{U}(\mathbf{x},\bm{\theta})$ that can be classically simulated. 

More importantly, our method is based on a quantum learning framework in which the parameters, the quantum gates and the circuit topology are optimized. 
The result is an algorithm that maximizes the classification accuracy and generalizability of the quantum model while minimizing the model complexity.

The structural optimization of the quantum classifier, which is accomplished through state-of-the-art evolutionary optimization algorithms, has several advantages. First, the process makes optimal use of an expensive resource: the quantum gates. By decreasing the size of the quantum circuit to the minimal required expression, we can run these algorithms in near-term hardware with the fewest number of errors. Second, and most importantly, the structural optimization identifies the circuits with the lowest expressive power that can reproduce the underlying model. This simplification allows us to prevent or reduce the problem of exponentially vanishing gradients or \textit{barren plateaus}~\cite{barren18}, which have been observed with other quantum machine learning strategies. Finally, the circuit optimization often reveals a simple structure that requires little or no entanglement~\cite{2021saj}. These circuits are referred to as quantum-inspired solutions because they can be implemented in classical hardware without the use of quantum computers.

\section{Evolutionary Quantum Classifiers}
\label{sec:EQC}

The use of evolutionary techniques allows the space of possible solutions to be explored to determine the best QSVM. Moreover, evolutionary systems can avoid falling into local minima through the use of mutations ~\cite{zhao2010ant}, which is a limiting factor of the training phase in other quantum ML models.

Since we consider two objectives during the optimization process, we use the Pareto front ~\cite{pareto1,pareto2}, which is updated during each generation, to satisfy both objectives, namely, achieving the best accuracy on the classification of the dataset and generating the smallest possible circuit.

\subsection{Quantum Classifier Encoding}
\label{sec:QCL}
\begin{figure*}
    \centering
    \includegraphics[width=0.8\textwidth]{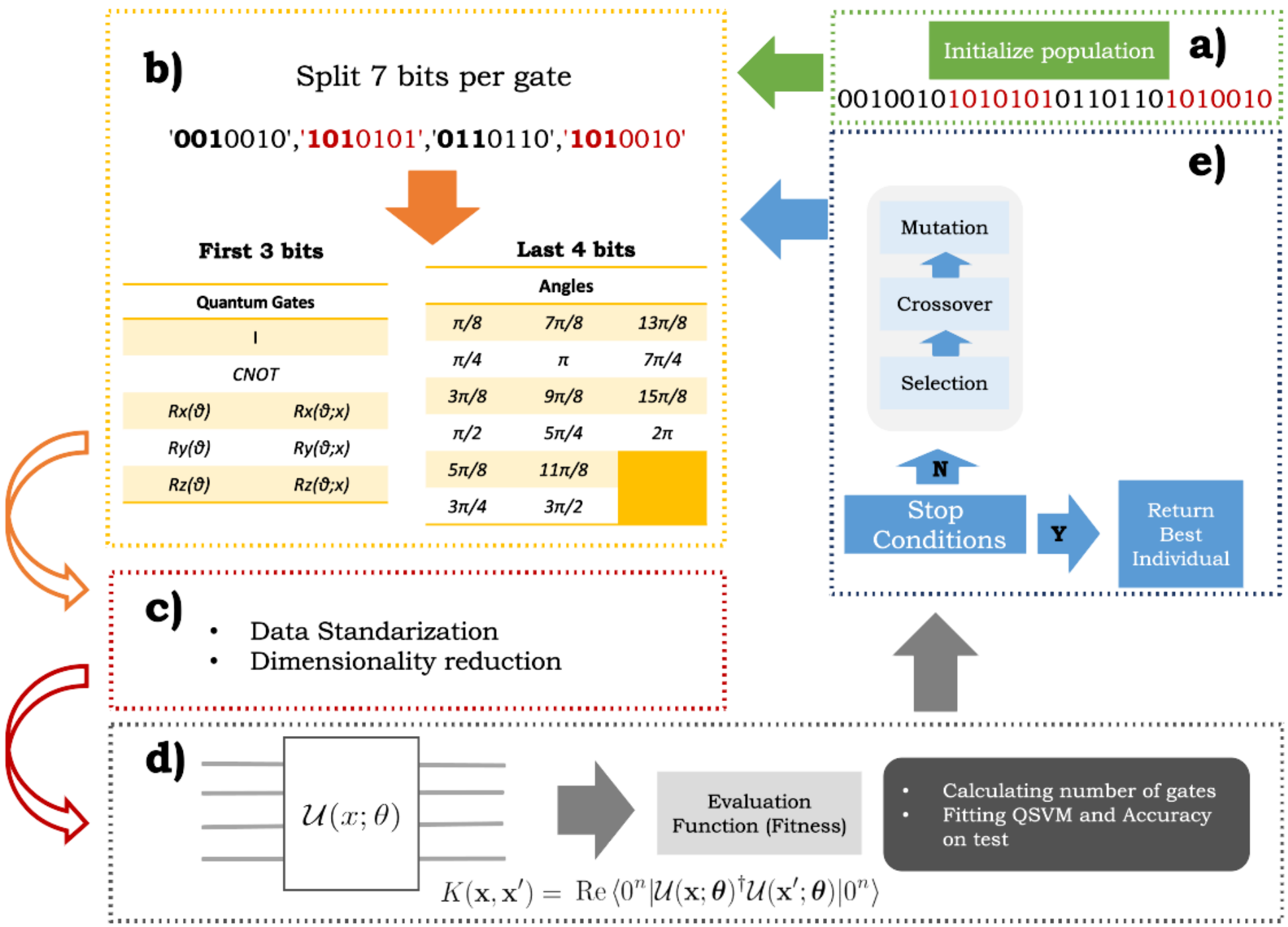}
    \caption{Evolutionary quantum classifiers. a) The population is initialized as random binary strings. b) Each gate is composed of 7 bits. The first three bits correspond to the gate type, while the remaining four bits correspond to the angle if necessary. c) Standardization and image dimensionality reduction method. d) We use the decoding individual to build the circuit and QSVM. We use the fitness function to search for the best accuracy on the test set and the least complexity. e) We use typical genetic operators to create the next generations until the stop condition is reached. }
    \label{fig:ga}
\end{figure*}

We use binary code strings as individuals as it can be seen in figure \ref{fig:ga}. The length of these strings is defined by the maximum number of input qubits (M) and the maximum number of layers (N) in the quantum circuit.

\begin{figure}[!t]
\centering
\includegraphics[width=3.3 in]{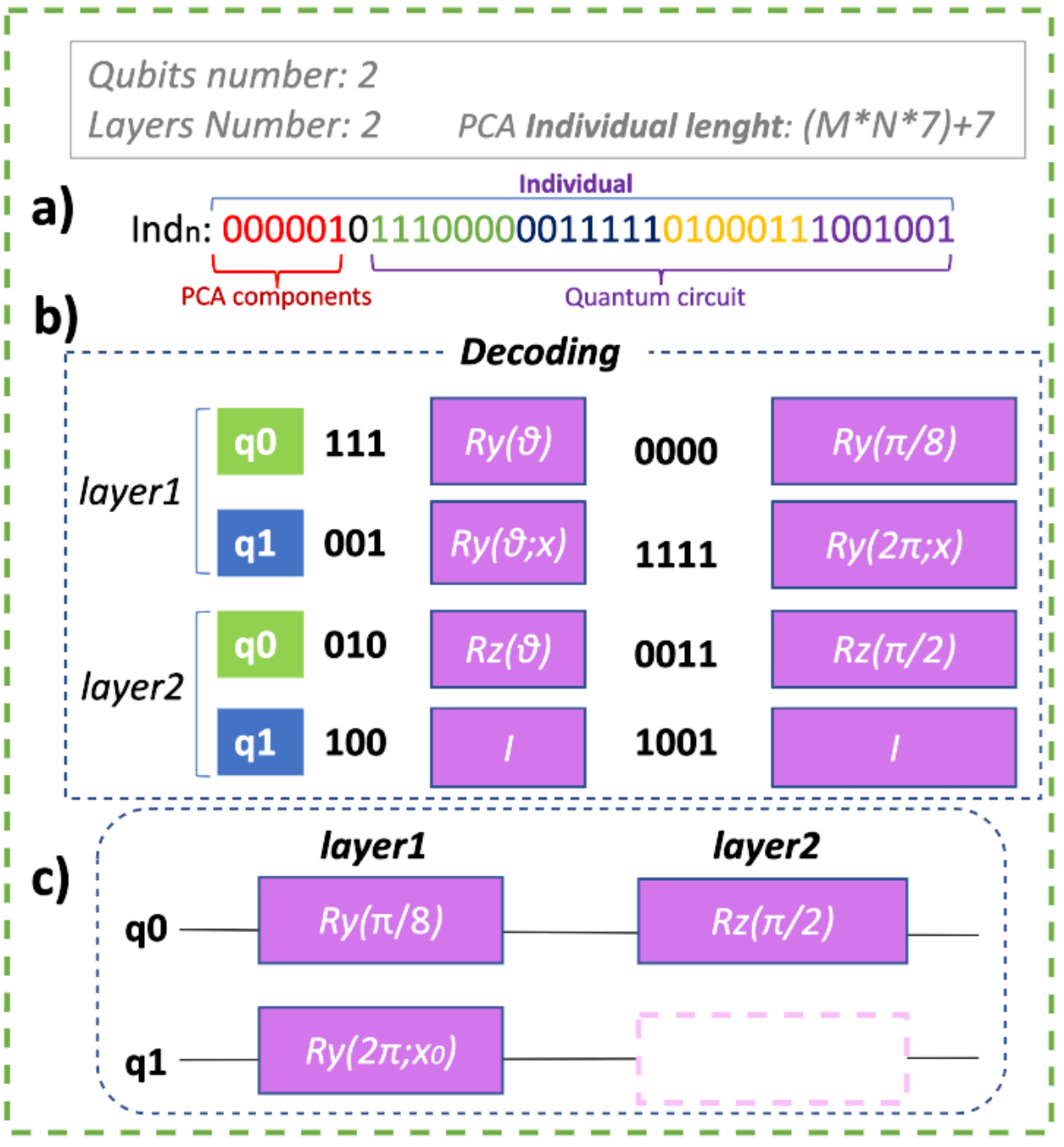}
\caption{Quantum circuit encoding. a) The individual is created based on the number of qubits and number of layers. Each gate is composed of seven bits. The first six bits in the individual are used as the number of components in the PCA approach. b) Coding phase. The first three bits correspond to the gate types, while the last four bits correspond to the angle. c) We assemble the quantum circuit, taking into account the predefined number of qubits and layers. The identity operator (I) allows us to reduce the size of the ansatz during the evolution.}
\label{fig_4}
\end{figure}

In this matrix, the quantum gates are positioned per layers in a vertical way as it can be seen in figure~\ref{fig_4}, being the maximum number of gates per layer the number of qubits (M), and the maximum number of layers (N) the defined by the user. The genes of a gate are defined by seven bits, namely, $g_0, g_1, g_2, g_3, g_4, g_5$, and $g_6$. The first three bits, $g_0, g_1$, and $g_2$, correspond to eight gate types, while the remaining four bits, $g_3, g_4, g_5$, and $g_6$, correspond to the angle introduced in the rotation, as shown in Figure \ref{fig:ga}b.

In this approach, local quantum gates are precoded to allow rotations about the three axes $R_x, R_y, R_z$. Furthermore, this approach includes entangling operators (CNOTs), which correlate consecutive qubits and identity operators, allowing the reduction of the dimension of the classifier during the evolution, as shown in Figure \ref{fig_4}c.

To increase the accuracy of the system and allow a higher degree of freedom during rotations, we define two types of rotations. In the first approach, the classical variables are embedded in the operators; in this case, the rotations are parameterized by the input vector as $R_x(\theta;x)$, $R_y(\theta;x)$, $R_z(\theta;x)$, where $R_\beta(\theta_i x_k)=\exp(-i\theta_ix_k\sigma^\beta)$. On the other hand, we encode fixed rotations that do not depend on the input data, $R_x(\theta)$, $R_y(\theta)$, and $R_z(\theta)$, with the expression $R_\gamma(\theta_i)=\exp(-i\theta_i\sigma^\gamma)$. These angles are a uniform sampling in the interval {$[\pi/8$, $2 \pi$]} according to the $(n \pi/8)$ rule, with $n$ in $[1, 16]$ due to the use of $2^4$ in the angle encoding. The gate selection is motivated to be a universal set. Since the encoded angles in our quantum gates can also produce H, T and S gates with fixed angle rotations (see Section~\ref{subsec:qugates}).

Based on this encoding, the general length of the individual's string is calculated as $MxNx7$. Since we have enconded gates which do not depend on input variables of the dataset ---fixed rotations $R_\gamma(\theta_i)$, these gates can be replaced by gates with inputs, which leads to a selection of variables in the provided quantum models. Furthermore, some variables can be replaced by identity operators, reducing the size of the circuit. Thus, the best classifier in the evolution has the optimal quantum gate combination for the ad-hoc classification task.

\subsection{Dynamic Fitness Function}

The objectives are to maximize the accuracy of the test set while minimizing the size of the quantum circuit. First, we divide the dataset into two subsets, the training set and the test set, to minimize data leakage. We stratify the datasets based on the target. The data are standardized to [-1,1], due that some dimensionality reduction methods because large differences in the ranges of the variables can lead to biased results that are influenced more by variables with larger values  (See Section ~\ref{sec:DR}).

After we prepare our data, we decode the individual in the quantum circuit based on Section \ref{sec:QCL}. The individual is then used as a quantum feature map in the QSVM, where the input data are embedded. As shown in Figure \ref{fig:fig_second_case}1, the model is fitted with the training data. Then, we use the model to predict the test set, which includes data not previously seen by the model, indicating the generalizability and robustness of the model. Finally, the success rate of the classifier can be calculated.

In addition, we can evaluate the second fitness objective: the number of gates in each individual. We calculate the success rate according to Equation \ref{eq:cf}. We associate the \textit{complexity} of the circuit with the number of gates. Due to the higher computational cost of the entanglement gates, as well as the fact that we want quantum-inspired solutions, these gates are weighted by two, penalizing their appearance. Additionally, because the identity operators add 0 to the equation, as shown in Figure \ref{fig_4}, we can use these operators to reduce the size of the solution. Because the solution has fewer entangling operators, we can program the classifier in a classical machine as a simple matrix multiplication operation.

\begin{equation}
  \label{eq:cf}
\text{Complexity (C)} = \frac{N_\text{local} + 2 N_\text{CNOT} + 0 N_\text{I} }{N_\text{qubits}}
\end{equation}

Related to the fitness objetives, when we assign a high weight to the accuracy objective, the genetic variability of the evolution can be reduced. On the other hand, when we assign a high weight to the size reduction objetive, it is possible that we lose too much information in the classifier, and the optimal ratio may not be obtained.

Therefore, we propose a \textit{dynamic fitness function} in which the \textit{complexity} (C) is a function of the increase in the squared accuracy that is formulated as follows:

\begin{equation}
\text{Objective balance \((O_B)\)} = \text{C} + \text{C} * \text{accuracy$_{test}$}^2.
\label{eq:weights}
\end{equation}

As a result, the two metrics can be balanced, allowing us to satisfy both objectives.

\subsection{Genetic Operators}
We use a multiobjective genetic algorithm to identify the best individual. A NSGA-II selection algorithm was used to maximize the accuracy while minimizing the size of the quantum circuit \cite{nsga}. Since the goal of this technique is to determine the best individual that satisfies both objectives, using the Pareto front to store and update the solutions throughout the evolution. 

The \textit{flipbit} mutation operator is selected due to the binary coding of the individuals, which include the probabilities of both individual mutations $p^{ind}$ and gene mutations $p^{gen}$. 
The \textit{two-point crossover} operator is used in this work. This technique involves choosing two random points from the parents and exchanging their genetic information with a probability of $p^{cross}$.

Based on this experiment, we obtain the best classifier. Then, we use the $\mu + \lambda$ strategy, which creates competition between the parents and their offspring, ensuring that the best individuals survive throughout the generations and increasing the optimal characteristics of the system. A change in a gene results a completely different circuit structure, including the gate type, number of embedded variables and angles. Furthermore, we use a dynamic fitness function, which allows the two objectives to be balanced while ensuring that genetic diversity is not lost along the evolution.

The experiments are repeated with different crossover hyperparameters and mutation probabilities to determine the optimal hyperparameters in both cases. As a result, we define $\mu$ and $\lambda$ as 50 and 20 individuals, respectively, where $\mu$ corresponds to the number of individuals selected for the next generation and $\lambda$ corresponds to the number of children produced during each generation \cite{deap}. We limited the evolution to 2000 generations because our aim was to compare the systems under the same conditions, and CAE system converge quickly, with the best individual found in less than 1000 generations.

\section{Use Cases}
\label{sec:DS}
\begin{figure}[!t]
\centering
\includegraphics[width=3.3 in]{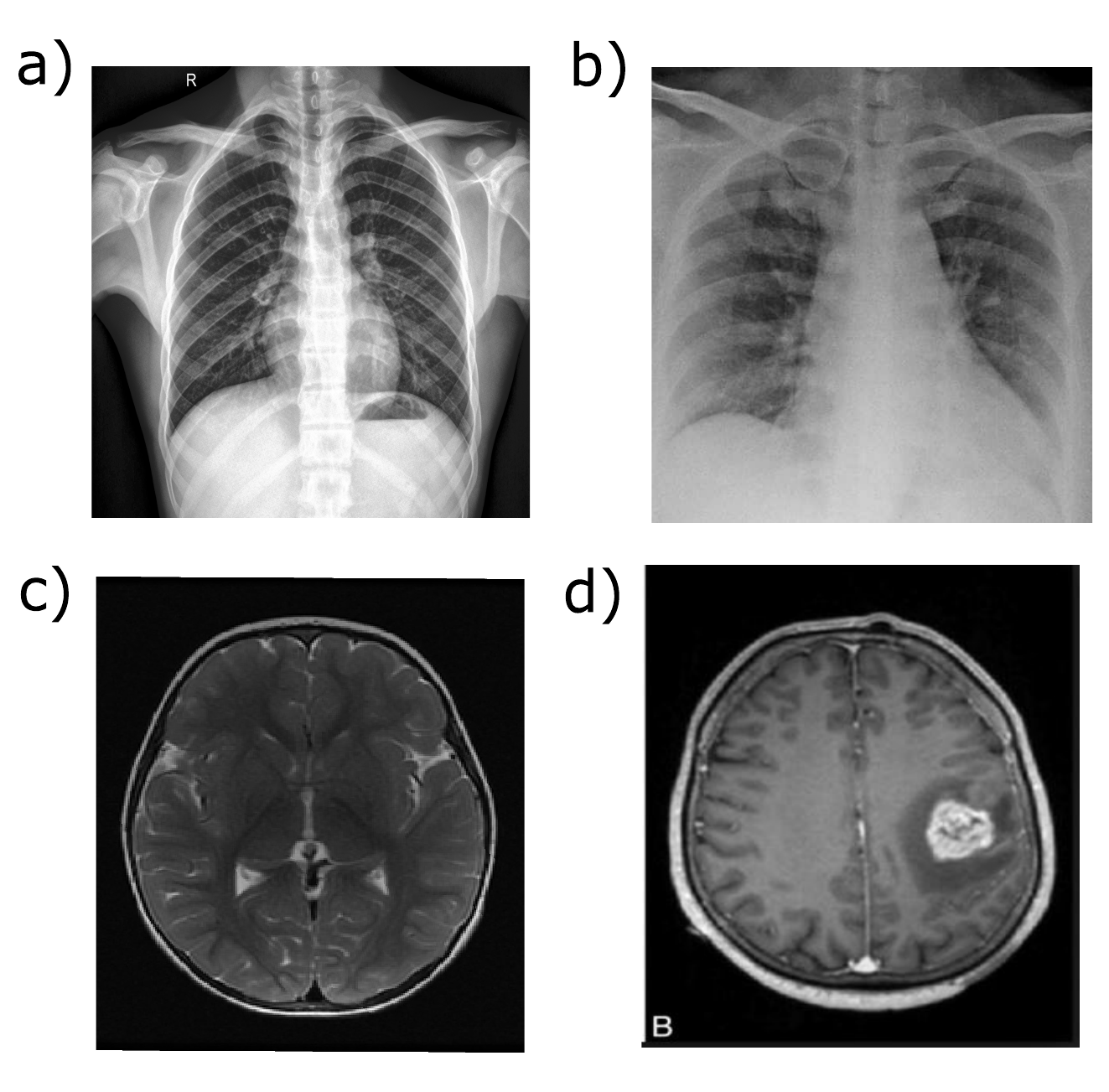}
\caption{Samples from the datasets used in the experiments, displaying pneumonia and a brain tumor. a and c show images with normal diagnoses from the two datasets ~\cite{DS:Brain}, while b and d represent images of COVID-19 and a brain structure anomaly, respectively ~\cite{DS:COVID19}.}
\label{fig_1}
\end{figure}

In the present study, we used two types of medical images to evaluate the proposed technique. We used a dataset that includes X-ray images of lungs from patients affected by SARS-CoV-2 (COVID-19) and healthy patients \cite{DS:COVID19}. The dataset includes 227 images. Two categories, were extracted from the image set by implementing a supervised algorithm to identify the classes, as shown in Figure \ref{fig_1}(a, b).

We also used a dataset with 253 images. The dataset included magnetic resonance imaging (MRI) images, some displaying brain tumors and some with healthy brain structures \cite{DS:Brain}. 
In this case, our aim was to identify abnormalities in the brain structure. Figure \ref{fig_1}(c, d) shows both classes, patients with positive and negative diagnoses.

Several simulations were performed with these datasets to determine the optimal probabilities for the genetic operators during the evolution of each dataset. For the COVID-19 dataset, the probabilities of $p^{ind}$ and $p^{cross}$ were chosen to be 0.4 and 0.6, respectively, and $p^{gen}$ was set to 0.3. For the brain tumor dataset, we defined $p^{ind}$ and $p^{cross}$ as 0.3 and 0.7, respectively, and $p^{gen}$ was set to 0.4.

Due to the high dimensionality of the images in the datasets, dimensionality reduction techniques were used to ensure that the information could be handled by a quantum circuit. For both cases, the same preprocessing techniques, which are described in Section ~\ref{sec:DR}, were used.

\section{Dimensionality Reduction}
\label{sec:DR}

The datasets contain high-dimensional data $D$ ---images. A widely used methodology for image classification problems is to perform dimensionality reduction before inputting the data into the classifier. These techniques allow the most relevant features of the dataset to be identified to distinguish each category, summarizing the global information in $d$ dimensions, where $d << D$. In quantum machine learning, dimensionality reduction is an essential step because it allows the image input data to be embedded into the quantum circuit.

In this work, we use two techniques with the quantum classifier and analyze their behaviors. Both methods are formally described in the following subsections. It is important to ensure that the classification power is due to the quantum circuit and that the dimensionality reduction technique does not pre-classify the data. Therefore, we use different strategies for each technique.

\subsection{PCA: Principal Component Analysis}
Principal component analysis (PCA) is a nontrainable technique that allows us to project the original data to lower dimensional spaces. Given a $Z$ dimensional dataset, this technique allows us to determine $Y$ underlying variables ($Y<Z$) that maintain a certain percentage of the variance in the original set, thus condensing the information to $Y$ dimensions or \textit{components}. The covariance matrix is used to identify the redundant information of the possible correlated variables in the $Z$ dimensional dataset.

These components are linear combinations of the original variables with associated weights, and the components are perpendicular to each other. In this method, the components are the eigenvectors, and they are ordered according to their eigenvalue in decreasing order, e.g., the first component of a dimensionality reduction with PCA is the eigenvector with the greatest eigenvalue.

In the PCA method, the singular value decomposition (SVD)~\cite{halko2010finding, sadek2012svd,szlam2014implementation,martinsson2011randomized} solver is a randomized algorithm due to its high computational efficiency and its ability to handle sparse matrices. In this method, which is not a deep learning method, the images are homogenized to a size of 250 x 250. In this case, the number of components required for the implementation is encoded in the individuals; thus, the optimal number of components can be chosen according to the system.

\subsection{CAE: Convolutional Autoencoders}

In this work, we use a convolutional autoencoder (CAE) network as a second method to understand the behavior of the technique after information is extracted by a neural network that is not a large deep learning network.

The main objective of autoencoder neural networks is to reproduce the inputs as accurately as possible in the outputs of the network, maintaining the same dimensions. The autoencoder neural network is an unsupervised learning method that allows information to be summarized in a vector, which belong to the so called \textit{latent space}.

The network architecture includes an \textit{encoding} component, in which the dimensionality is reduced by fully connecting the network with the latent space. 
At this point, the input image is reproduced using the latent space data in such a way that the cost function is as small as possible. This phase is called the \textit{decoder}. Since the data in this study are images, we can use simple convolutional layers as encoders and decoders to perform dimensionality reduction in the latent space ~\cite{zhang2018better}.

To accelerate the training process and reduce the size of the network without compromising the output, the size of the images is reduced. The images used in the CAE approach are homogenized to 28x28 so that the dimensionality reduction technique does not utilize a large deep learning structure and thus classify the data itself; instead, the technique uses simple information extraction. Thus, the encoding part of the CAE is composed of a 28x28 input layer and two simple convolutional layers with maxpooling up to 64 dimensions in the latent space. Then, a decoding component with the same structure is implemented to produce as similar an output image as possible. The network is trained over 250 epochs. The Adam optimizer~\cite{zhang2018improved} with a binary cross-entropy loss function is used in the implementation.

\subsection{Dimensionality Reduction Encoding}

Due to the size of the images used in this study, we apply dimensionality reduction methods. 

\begin{figure*}[!t]
\centering
\subfloat[]{\includegraphics[width=3.3in]{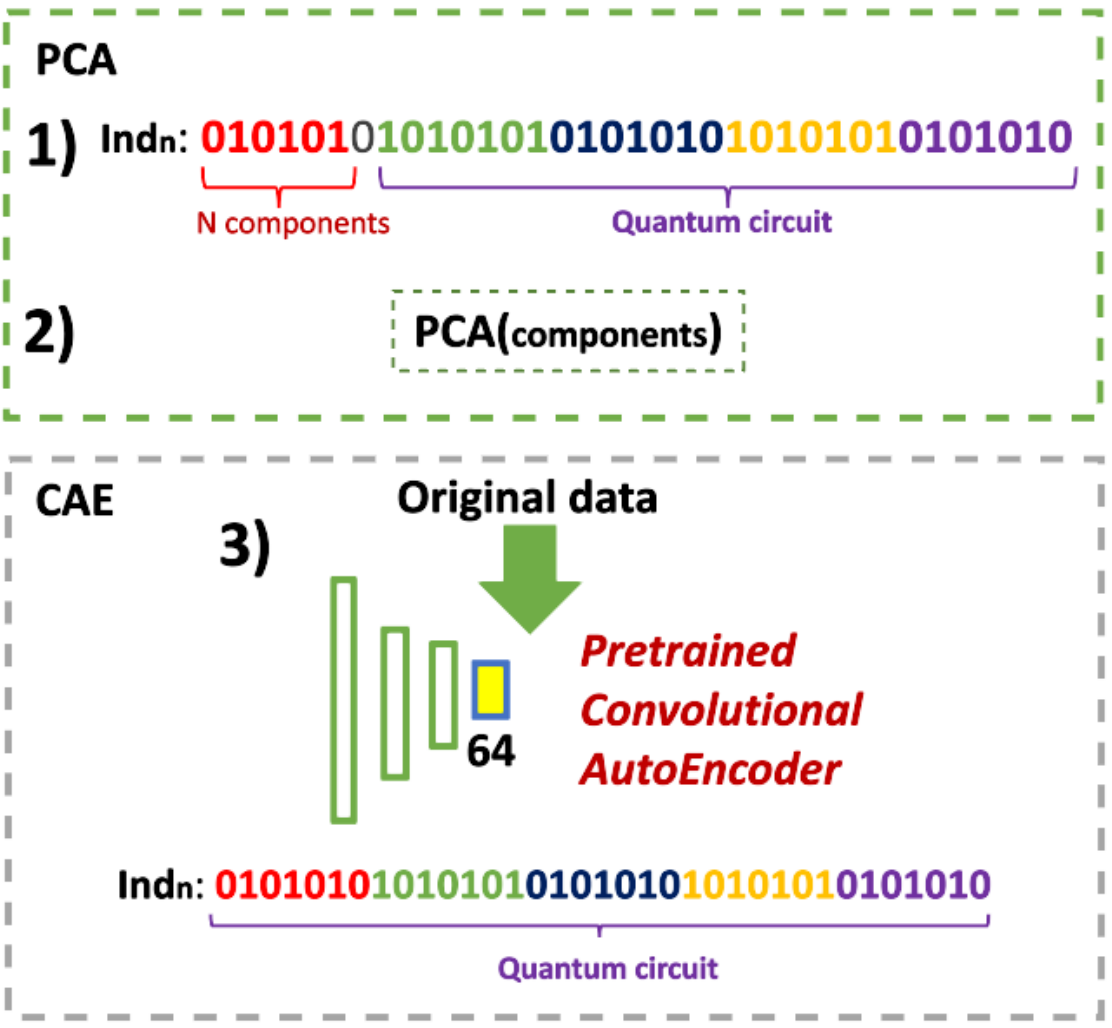}%
\label{fig:fig_first_case}}
\hfill
\subfloat[]{\includegraphics[width=3.3in]{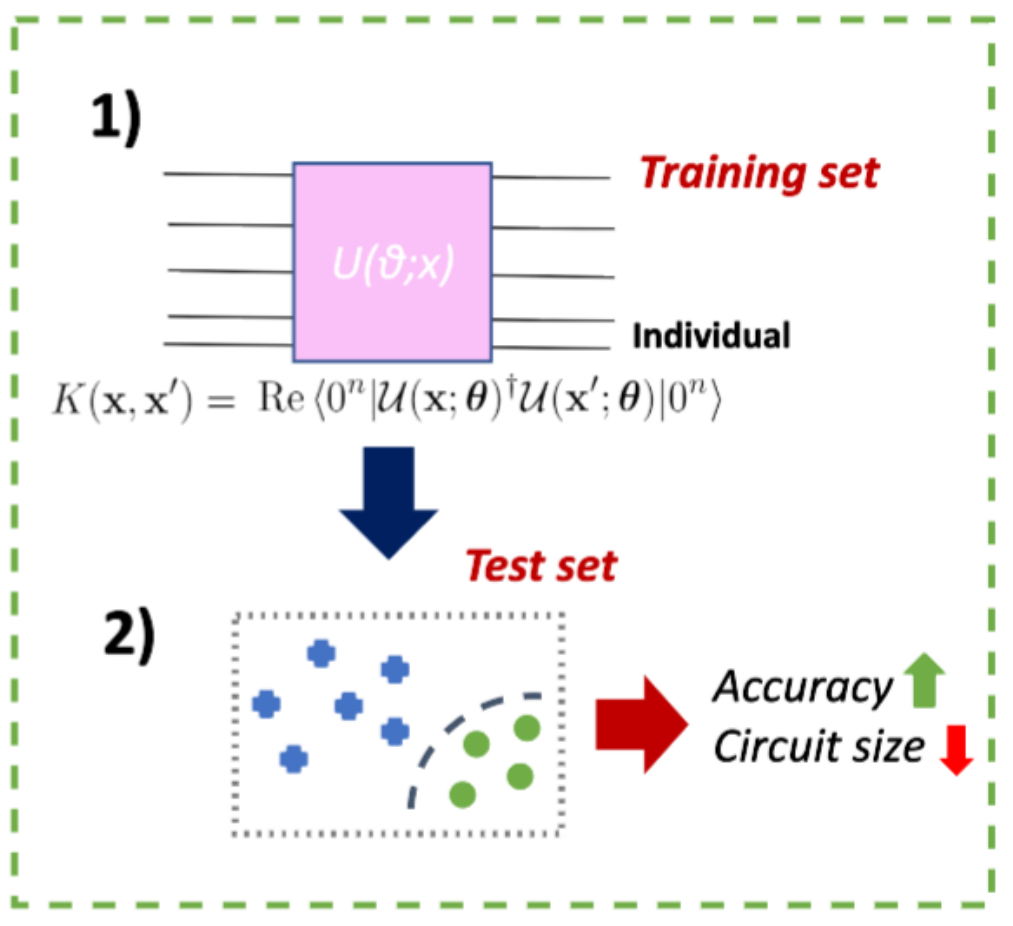}%
\label{fig:fig_second_case}}
\caption{Fitness function. a) Dimensionality reduction methods. Preprocessing. We evaluate two preprocessing methods. In the PCA approach, we use six bits of the individual to reduce the dimensionality, as shown in 1 and 2. On the other hand, in CAE approach, we apply a pretrained model with a fixed a 64-dimensional output so that the whole bit string belongs to the individual. b) Training. After preprocessing the data, we build a circuit based on the individual coding, fit the parameters and test with unseen data. Finally, we calculate the success rate and weighted circuit size.}
\label{fig_sim}
\end{figure*}

One of the proposed approaches is to use a nontrainable technique for dimensionality reduction, such as PCA. For this approach, we add the number of components in the individual in order to make the system self-tuning and use the best number of components for the circuit as can be seen in figure ~\ref{fig_4}. For this reason, and also to maintain the structure and proportions in quantum circuit encoding, we add 7 bits to the individual. In this case, We use only six bits to ensure a fair reduction and comparison with the other approximation methods; however, more bits could be used in this encoding method. The number of features is limited to 6 bits, resulting in a maximum of 64 dimensions, as $2^6$ bits. Thus, the string length is calculated as $7+MxNx7$. 

Thus, we can optimize our method to determine the optimal number of PCA components to be used in the best topology of coded gates and parameters. Due to the behavior of the PCA technique, which orders the components according to their variance contribution, the variables are input into the model in a linear manner ---straightforward.

In the CAE approach, we pretrain a convolutional autoencoder neural network to perform 64-dimensional extraction, which is equivalent to six bits. This method can be considered to be a form of \textit{transfer learning}, allowing us to determine how our method behaves. In the CAE training, we apply the encoding part of the network to the input data and obtain the same latent space in 64 dimensions, as shown in Figure \ref{fig:fig_first_case}3.

\section{Training Process}
\label{sec:AL}

In this section, we explain the steps in the \textit{evolutionary quantum-inspired machine learning} (EQI-ML) technique, taking into account the PCA and CAE dimensionality reduction processes, which are described in Section ~\ref{sec:DR}.

\subsection{EQI-ML Algorithm - PCA Approach}
\textbf{Step 1.} We define the maximum size of the circuit based on the number of qubits (M) and the number of layers (N). We add seven bits to use as the number of components. Since we limit the algorithm to 64 components, six bits are used.
\textbf{Step 2.} Vectors are created based on the images, and the training and testing data are divided and normalized. 
\textbf{Step 3.} The individual enters the fitness function and is separated according to the circuit (MxNx7) and the number of components (7 bits). Then, first six bits are taken and passed as an integer, which is the number of components to be added in the PCA method. \textbf{Step 4.} The PCA transformation is applied to the training and testing data. The individual is decoded to produce a quantum feature map in which the dimensions derived according to the embedded PCA features. We fit the QSVM with the training set. 
\textbf{Step 5.} The prediction on the test data is obtained, and the size of the circuit is calculated based on the weights; these features are both objectives of the genetic algorithm. After this process, the optimal individuals are stored in the Pareto front, and the genetic operators are applied to create the next generation. 
\textbf{Step 6.} Iterate beginning from \textbf{Step 3} until the algorithm converges or the stop conditions of the genetic algorithm are reached.

\subsection{EQI-ML Algorithm - CAE Approach}
\textbf{Step 1.} A small CAE neural network that does not require deep learning is implemented to extract information, reducing the \textit{encoding} part of the network, which has an output of 64 dimensions. 
\textbf{Step 2.} The maximum circuit size based on the number of input qubits (M) and the number of layers (N) is determined with the expression (MxNx7). 
\textbf{Step 3.} We divide the dataset into training and test sets. Both datasets are normalized, and the pretrained model is applied to the images, obtaining 64 fixed dimensions. 
\textbf{Step 4.} The individual enters the fitness function and is decoded, generating a quantum feature map in which the input CAE variables are embedded in the quantum gates to fit the QSVM. 
\textbf{Step 5.} The prediction on the test set is obtained, and the size of the circuit is calculated based on the weights; both of these features are objectives of the genetic algorithm. After this process, the best individuals are stored in the Pareto front, and genetic operators are applied to create the next generation. 
\textbf{Step. 6} We iterate the processes starting from \textbf{Step 4} until the algorithm converges or the defined stop conditions are reached.

\section{Results and Discussion}
\label{sec:Reesults}

We use the same preprocessing techniques for both cases. We first divide the dataset into training and testing sets with proportions of 75\% and 25\%, respectively, to prevent data leakage effects. The data are standardized to [-1,1] due that some dimensionality reduction methods require it for a good performance (See Section ~\ref{sec:DR}).

Both experiments use three implementations: PCA($N$) + QSVM, CAE(64) + QSVM, and PCA(64) + MLP. Our goal is to determine which dimensionality reduction method provides the most power to the quantum system by comparing the methods. We also compare the results with a classical nonlinear system.

Since we use six bits for PCA, which results in a maximum of 64 dimensions, we predefine a circuit size for all tests that can embed all the variables that depend on the input parameters if necessary. Therefore, the maximum number of qubits is set to six, and the maximum number of layers is set to eleven, thus 66 gates as maximum in the circuit. $N$ is the number of components applied, which is defined by each individual. In the CAE system, the circuit is fixed to 64 input dimensions according to the training output of the convolutional autoencoder neural network.

For the classical comparison, a multilayer perceptron (MLP) with nonlinear transformations, an ReLU layer, and one hidden layer was used. The images are homogenized to 250x250 in these models. To ensure that the number of parameters learned by the network was fair in the comparison, the number of neurons in the hidden layer was chosen based on the number of qubits in the circuit. Thus, this nonlinear classifier is composed of an input layer with 64 dimensions based on the output of the PCA application under the same conditions as in the quantum system, a hidden layer with six neurons and an output layer with two neurons based on the binary classification task.

\subsubsection{\textbf{COVID-19}}
\begin{figure*}
\centerline{\includegraphics[width=\textwidth]{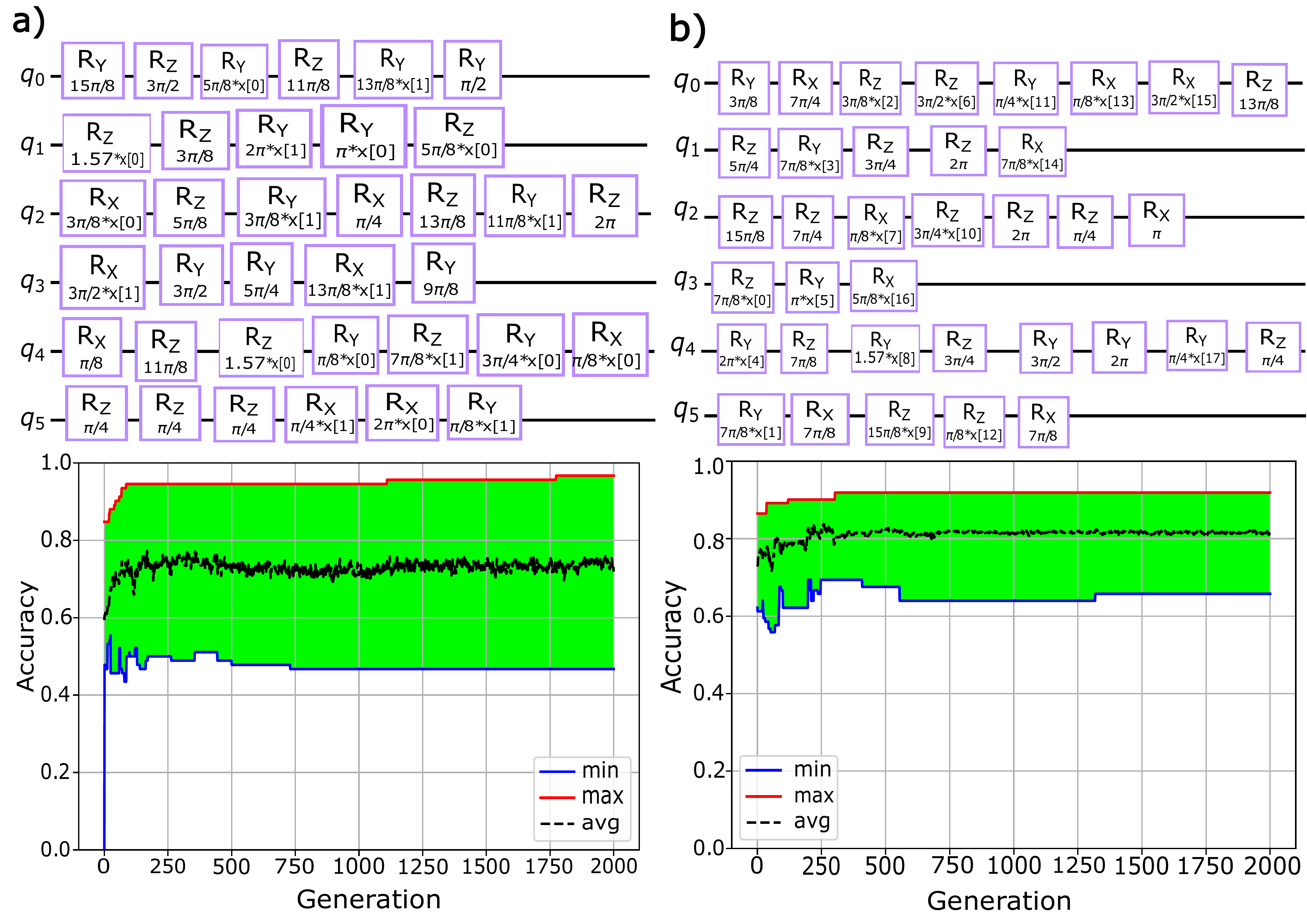}}
\caption{We observe the best circuits in the evolution of both hybrid quantum systems for the COVID-19 dataset. a) The best PCA+QSVM system uses 2 components. We can observe the best individual in the evolution. b) The best circuit in the evolution of the CAE+QSVM system for the COVID-19 dataset has an input of 64 dimensions but only 18 were used by the model.}
\label{fig:fig_sim_covid}
\end{figure*}

We apply both hybrid classical-quantum systems to the classification task to create a classifier for unseen data. 
Figure \ref{fig:fig_sim_covid} shows the results obtained after finishing the evolution. We observe that the circuits do not have entanglement gates. When the PCA+QSVM implementation is applied, the number of components depends on the individual. Two components are used to assemble this circuit, as shown in Figure \ref{fig:fig_sim_covid}a. The best individual achieved in the evolution with this technique has an accuracy of 0.967.

On the other hand, when the CAE+QSVM approach is applied, the best solution is a classifier composed of local gates and no entanglement gates. In the fitness function, 64 dimensions are passed to the circuit from the pretrained model. As shown in Figure \ref{fig:fig_sim_covid}b, in the distribution of the final circuit, the input vector has 18 variables, with the remainder of the gates having fixed rotations or identities. The best accuracy achieved in the evolution was 0.919.

A nonlinear MLP with ReLU transformations was used as a basis. The hyperparameters were selected after several tests to avoid overfitting effects and to determine the best learning rate. The learning rate was 0.01 for 100 epochs because the nonlinear MLP does not improve its rate with more epochs. The maximum accuracy achieved on the test set was 0.945. The cost function used the binary cross entropy with an Adam optimizer due to the binary nature of the task.

\subsubsection{\textbf{Brain Tumor}}
\begin{figure*}
\centerline{\includegraphics[width=\textwidth]{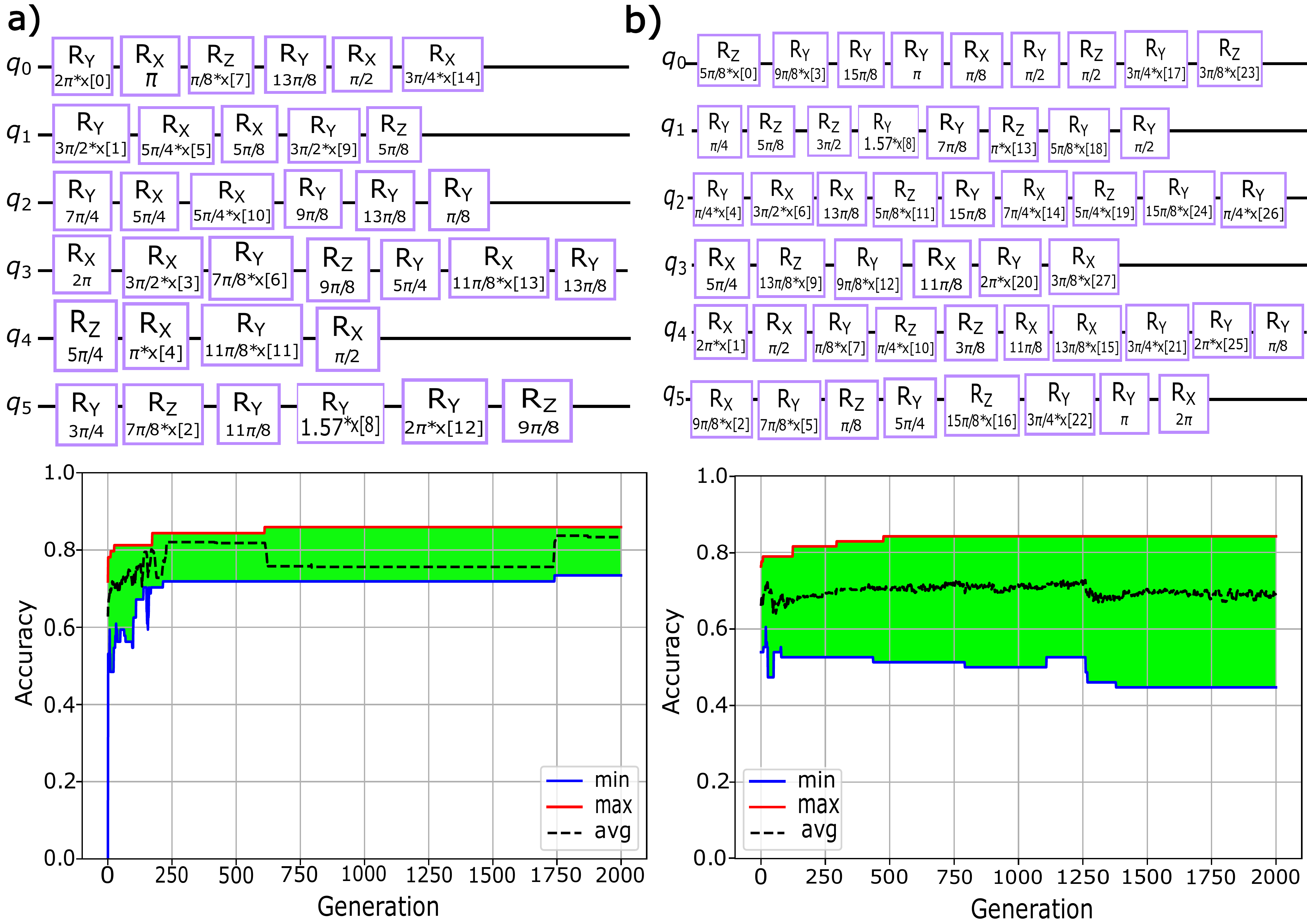}}
\caption{We observe the best circuits in the evolutions of both hybrid quantum systems for the brain tumor dataset. a) The PCA+QSVM system uses 45 components. We observe the best individual that was automatically generated during the evolution. b) The best circuit for the CAE+QSVM system and brain tumor dataset in terms of size and accuracy uses 27 dimensions of the 64 dimensions vector.}
\label{fig:fig_sim_brain}
\end{figure*}

The results obtained with both quantum systems are shown in Figure \ref{fig:fig_sim_brain}. 
The best PQC obtained with the PCA+QSVM system is shown in Figure \ref{fig:fig_sim_brain}a. The best individual has 45 components; however, only 15 components appear in the solution, while the remaining components have fixed rotations. The maximum accuracy achieved in the evolution is 0.859.

On the other hand, the CAE+QSVM system has a 64-dimensional input, but only 27 dimensions are used in the system, with the rest of the operators being noninput-dependent or identities, thus allowing the variables to be selected. The best quantum circuit, which is shown in Figure \ref{fig:fig_sim_brain}b, achieves a success rate of 0.846 on the test data, as seen in its evolution.

Both approaches can obtain quantum circuits with no correlations.

A nonlinear MLP with ReLU transformations was used as a basis to understand the segmentation power of the dimensionality reduction methods to ensure that the classification power is due to the PQC. The hyperparameters were selected after several tests to avoid overfitting effects and to determine the best learning rate. The learning rate was 0.001 for 100 epochs. The cost function used the binary cross entropy with an Adam optimizer due to the binary nature of the task. The maximum accuracy achieved on the test set was 0.684.

\begin{table}[!t]
\caption{Accuracy of the quantum and classical models.\label{tab:table1}}
\centering
\begin{tabular}{|l|l|l|l|l|l|l}
\hline
Use Case & Models &Accuracy\\
\hline
\multirow{3}{*}{COVID-19}
& \textbf{PCA(2) + QSVM}& \textcolor{teal}{\textbf{0.967}}\\
 &  \textbf{CAE(64) + QSVM}&0.919\\
 &  \textbf{PCA(64) + MLP}& 0.945\\
 \hline
\multirow{3}{*}{Brain tumor}
 & \textbf{PCA(45) + QSVM}& \textcolor{teal}{\textbf{0.859}}\\
 & \textbf{CAE(64) + QSVM}&0.846 \\
 & \textbf{PCA(64) + MLP}& 0.684\\
 \hline
\end{tabular}
\end{table}

Table \ref{tab:table1} shows a comparison in terms of the test set accuracy based on the unseen data in the model fitting, the use case and the method used. In this table, we observe in color blue the models that provided the best results for each dataset. Numerically, we observe that the PCA+QSVM system appears to obtain better results than the CAE+QSVM system in both use cases. One of the reasons for this result may be the existence of the number of components within the individual, which allows the dimensionality reduction method to be automatically optimized when developing the circuit, while in a pretrained system, the number of dimensions is always fixed. In both use cases we can see that the accuracy is higher using QML than classical methods, so we can conclude that there is a quantum advantage.

We ensure that the PCA transformation cannot provide a direct match with the target itself, as the MLP does not improve the predictive capability of the QSVM approach. Therefore, it can be concluded that the circuit results in the predictive capability of the system.

Interestingly, in the first generations of the evolution, we observe that the generated circuits have large sizes and high complexities in terms of the number of entangling operator gates that disappear as the individuals evolve. The final solutions have no correlations. As demonstrated in Ref. \cite{2021saj}, each qubit works as an independent kernel. This technique searches for the best genes such that the multiplication of all kernels $K(\mathbf{x},\mathbf{x}')=\prod_{i=1}^m K_i(\mathbf{x},\mathbf{x}')$ provides the most accurate prediction. Finally, we observe that the circuits are smaller than the predefined size, thus proving the good performance of the technique on both evolution objectives.

\section{Conclusions and Future Work}
\label{sec:conclusions}

In this work, we describe AutoQML, a technique for automatically generating and training quantum classifiers on grayscale images by using multiobjective genetic algorithms. This new hybrid method allows us to generate quantum classifiers with high generalizability because we determine the accuracy based on the test set rather than the training set, which ensures that generated classifiers are robust.

In terms of image processing, in quantum machine learning, it is important to address the issue of reducing the dimensionality of the input variables in the quantum circuits. In this study, we used two dimensionality reduction methods: a pretrained method (CAE) and a nontrainable method (PCA).  By including the number of components within the individual, we can optimize the input to the circuit, thus producing better results than pretrained systems, which include a fixed number of dimensions, since the variance in the PCA approach changes based on the number of components. 
The number of components was set to six bits to allow for a fair comparison of the different methods; however, this value can be increased to increase the number of components.

This system includes both gates with embedded input variables and noninput-dependent gates. This composition leads to a set of variables and rotations that can be optimally combined to produce the best prediction.

The application of weighting the gate typology allows circuits to be composed based on local rotations, not correlations, reducing the complexity of the generated quantum circuits. Thus, this approach can be considered an \textit{evolutionary quantum-inspired machine learning} technique, promoting a new research direction on the connection between classical and quantum algorithms. It may be interesting to study how this method performs in other application fields to determine possible improvements with respect to classical models.

Furthermore, we believe that the proposed technique can be used to develop quantum-inspired classifiers, which could provide some interpretability that is lost when using classical models, such as deep learning techniques, since our method uses a product of simple matrices. This can be an important point for quantum machine learning, since it differentiates quantum methods from classicals in \textit{ethical decisions} in artificial intelligence. Therefore, quantum-inspired ML could be used by companies to mitigate reputational risks.

{\appendix[Quantum Gate Encoding Tables]
\label{sec_ap}
The following tables ~\ref{tab:enco1}(a,b), include the gate type and its code, as well as the angles coded and the binary code for each parameter. As previously noted, the gate type uses 3-bit encoding, with $2^3$ combinations, while the angle gates use 4-bit encoding, with $2^4$ combinations.

\begin{figure}
        \centering
        \subfloat[]{%
           \begin{tabular}{|l|l|l|l|l|l|l}
            \hline
            Gate&Code\\
            \hline
            $R_x(\theta;x)$& 000\\
             \hline
             $R_y(\theta;x)$& 001\\
             \hline
              $R_z(\theta;x)$& 011\\
             \hline
             CNOT& 101\\
             \hline
             Identity & 100\\
             \hline
             $R_x(\theta)$& 110\\
             \hline
             $R_y(\theta)$& 111\\
             \hline
             $R_z(\theta)$& 010\\
             \hline
            \end{tabular}}\qquad
        \subfloat[]{%
            \begin{tabular}{|l|l|l|l|l|l|l}
            \hline
            Angle&Code\\
            \hline
            $\pi/8$& 0000\\
             \hline
             $\pi/4$& 0001\\
             \hline
             $3 \pi/8$& 0010\\
             \hline
            $\pi/2$& 0011\\
             \hline
            $5\pi/8$ & 0100\\
             \hline
             $3\pi/4$& 0101\\
             \hline
             $7\pi/8$& 0110\\
             \hline
            $\pi$& 0111\\
             \hline
            $9\pi/8$& 1000\\
             \hline
             $5\pi/4$& 1001\\
             \hline
             $11 \pi/8$& 1010\\
             \hline
            $3\pi/2$& 1011\\
             \hline
            $13\pi/8$ & 1100\\
             \hline
             $7\pi/4$& 1101\\
             \hline
             $15\pi/8$& 1110\\
             \hline
            $2 \pi$& 1111\\
             \hline
            \end{tabular}}
    \label{tab:enco1}
    \end{figure}
 }

\bibliographystyle{IEEEtran}
\bibliography{main}

\begin{IEEEbiographynophoto}{Sergio Altares-López} received his B.Sc. degree in agricultural engineering from the Technical University of Madrid, Madrid, Spain, in 2017, and his M.Sc. degree in business analytics and big data from the University of Alcala, Alcala, Spain, in 2018. He is currently a Ph.D. candidate in automation and robotics at the Technical University of Madrid, Madrid, Spain, a member of the Artificial Perception Group at CSIC, and a board member for the technical agricultural engineers association (CITAC), Spain. He has worked for multinational companies as a senior data scientist and analyst. His research interests include artificial intelligence, quantum machine learning and evolutionary techniques and their applications.
\end{IEEEbiographynophoto}

\begin{IEEEbiographynophoto}{Juan José García-Ripoll} received his Ph.D. in physics in 2001. He is a Senior Scientist working at the Institute of Fundamental Physics at the Spanish Research Council (CSIC). His main research interest is quantum technology, and he has contributed to the first designs of quantum simulations and quantum computers with ultracold atoms, trapped ions and superconducting quantum circuits. Currently, he combines this line of work with further research into quantum algorithms and quantum-inspired classical computation technologies. Among other duties, he coordinates the Quantum Information Group (Quinfog), the Spanish Network of Quantum Information and Quantum Technologies (RITCE) and the CSIC Platform on Quantum Technologies (QTEP).
\end{IEEEbiographynophoto}
\begin{IEEEbiographynophoto}{Angela Ribeiro} received her B.Sc. and Ph.D. degrees in physics from the Complutense University of Madrid. She is a Scientific Researcher at the Spanish National Research Council (CSIC) in the Centre for Automation and Robotics, (CSIC-UPM), where she is the leader of the Artificial Perception Group (GPA). Her research interests include artificial perception, pattern recognition, evolutionary algorithms, spatial knowledge representation, spatial reasoning for decision support systems, distributed systems, collective intelligence and multirobot systems. She has led several research projects on applications of the abovementioned techniques to address precision farming issues.
\end{IEEEbiographynophoto}
\end{document}